\documentclass{article}
\usepackage{amsmath,amssymb,amsbsy,amsthm}

\let\epsilon\varepsilon
\let\phi\varphi

\def\N{\mathbb N}

\def\E{{\operatorname{\bf{E}}}}

\def\X{\mathbf{X}}
\def\Z{\mathbf{Z}}
\def\N{\mathbb N}

\def\E{{\operatorname{\bf{E}}}}
\let\epsilon\varepsilon

\newtheorem*{lemma*}{Lemma}
\newtheorem{theorem}{Theorem}
\newtheorem{lemma}{Lemma}

\begin{document}

\title{Confidence Sets in Time--Series Filtering} 

\author{  Boris Ryabko and Daniil Ryabko  \\ \{boris, daniil\}@ ryabko.net }

\date{}

\maketitle


\begin{abstract}
The  problem of filtering of finite--alphabet stationary ergodic  time series is considered.
A method for constructing a   confidence set 
 for the (unknown) signal
is proposed, such that the resulting set has the following properties:
First, it includes the unknown signal with probability $\gamma$, where $\gamma$ is a parameter 
supplied to the filter.
Second, the size of the confidence sets grows exponentially with the rate that is
asymptotically equal to the conditional entropy of the signal given the data. 
Moreover, it is shown  that this 
rate is optimal. 
We also show that the described construction of the confidence set can be applied for the case
where the signal is corrupted by an erasure channel with unknown statistics. 
\end{abstract}


\section{Introduction}

The problem of estimating a discrete signal $X_1,\dots,X_t$  from a noisy version $Z_1,\dots,Z_t$  has 
attracted attention of many researchers due to its great importance for 
statistics, computer science, image processing, astronomy, biology,
cryptography, information theory and many other fields. 
The main  attention is usually focused on developing methods of estimation 
(denoising, or filtering) of the unknown signal, with the performance  measured  under
a given fidelity criterion; see \cite{Rissanen:00, RoosMR:09} and  references therein. 
Such an approach is  close in spirit to 
the problem  of point estimation in  statistics. 

An alternative approach, often considered in mathematical
statistics,  is that of constructing  confidence sets. 
That is, one tries to use the  data to construct a set 
that includes the unknown parameter (in our case, the signal) with 
a prescribed probability, while trying to keep the size of 
the set as small as possible (some classical examples of the use of this method in statistics can be found in, e.g.,  \cite{Kendall:61}).
Such a set is usually constructed as the set of most likely 
values of the parameter.

The reason why such an approach is of interest is as follows. 
In the presence of noise, the exact recovery of the signal is typically impossible,
and thus, in such cases, any of its estimates is necessarily imperfect. The choice of 
a particular estimate of the  signal out of many likely alternatives is largely arbitrary. 
Moreover, the optimal choice may depend on the specific application involved. 
The confidence--set approach effectively abstracts  from the problem of choosing
the ``best'' estimate, proposing, instead, a set of estimates. 
The performance of a method is then characterized by the size  of the confidence set (depending 
on the  confidence level).
This is the approach and the problems considered in this work.
We consider a model in which
 the
underlying noiseless signal  and the resulting corrupted (noisy) signal (and thus the channel) are assumed to
be stationary ergodic processes with finite alphabets. 
We mainly concentrate on the case where the probability  distributions 
of the noiseless  signal and the 
noisy channel are known.   
 (Obviously, in such a case the distribution of the corrupted signal is known, too.)
Besides, the case of a erasure channel with unknown 
distribution  is briefly mentioned, because in this case 
a conditional distribution of noiseless signal is known even though the distribution of the noise is unknown. 
The results that we obtain establish the optimal rate of growth (with respect to time, or to the length of the signal) 
of the size of the confidence set, as well as a method for constructing such a set. The optimal rate turns
out to be equal to the entropy of the signal given its noisy version.

Let us consider an example that illustrates our approach and exposes the notation.
Let the  signal be binary (with  the alphabet $\{0,1 \}$),  and suppose that it is 
 transmitted through a  memoryless binary erasure channel (e.g. \cite{Cover:06}). 
The binary erasure channel with erasure probability $\pi$ is defined as  a channel with binary input, 
ternary output (with the alphabet $\{0,1,*\}$), and the probability of erasure $\pi$. 
The channel replaces each input symbol 0 or 1 with the (output) symbol $*$  with probability $\pi$ (erasure), 
and places the input signal in the output otherwise (that is, with probability $1-\pi$).

 Suppose  that the noiseless sequence is generated by an i.i.d. source $P$ and
$P\{X_i = 0 \} = 0.9$, and let the erasure probability be any $\pi\in(0,1)$,
i.e. the erasure probability is unknown.
 Suppose that   the corrupted by noise  sequence is as  follows:
$$ Z_1 ... Z_4 = 0*1*  \, . $$ 
Then we have the following probability distribution for the lossless signal:
$$ P(\{X_1 ... X_4 = 0010 \} ) = 0.81, $$ $$  P(\{X_1 ... X_4 = 0110 \} ) = 0.09,  $$
$$ P(\{X_1 ... X_4 = 0011 \} ) = 0.09, $$ $$  P(\{X_1 ... X_4 = 0111 \} ) = 0.01 .  $$
If we take the confidence level $\gamma = 0.99$,  the  confidence set
will contain three following sequences: $\{ 0010, $ $  0110, $ $ 0011 \}$.

The goal of this paper is to describe a construction of confidence sets and to give an estimate of their size,
for the case when the  signal and noise  are stationary ergodic  processes with finite alphabets.
It is shown   that for any $\gamma \in (0,1)$ the size of the confidence set grows exponentially 
with the rate $h(X|Z)$, where $h(X|Z)$ is the limit (conditional) Shannon entropy.
This result is valid for the case when the probability distributions of noiseless signal and noise are known as well as for the
case when the probability distribution of the signal is known and the noise is described by a stationary erasure 
channel with memory whose probability distribution is unknown. 
 Moreover, we prove that the rate  
$h(X|Z)$ 
is  minimal, which  means that the suggested method of constructing   confidence sets is asymptotically optimal.

It is worth noting that the information theory is deeply connected with  statistics of time series and
signal processing; see, for example, \cite {Cover:06,Csiszar:04,Morvai:96, Rissanen:84, BRyabko:09,BRyabko:06a, Ryabko:101c, Ryabko:111c, Ryabko:103s} and 
\cite{Rissanen:00, RoosMR:09, Ordentlich:08}, correspondingly.
 In this paper a new connection of this kind is established: it is shown that the Shannon entropy determines
 the rate of growth of the size of the confidence set  for the signal, given its version corrupted by  stationary noise.

\section{The confidence sets and their properties} 

We consider the case where the signal $X = X_1,X_2,\dots$ and 
its noisy version $Z= Z_1,Z_2,\dots$  are described by 
stationary  ergodic processes with finite alphabets $\X$ and $\Z$ respectively. 
(There may be arbitrary long-range dependencies between the variables.)
 It is assumed that 
probability distributions of both processes are known, and, hence, the statistical structure of the noise
corrupting the signal  $X = X_1,X_2,\dots $ is known, too. 
Introduce the short-hand notation $X_{1..t}$ for $X_1,\dots,X_t$, and analogously for~$Z$.

Informally, for any  $\gamma \in (0,1) $ and  any sequence $Z_1,\dots,Z_t$ 
we define the confidence set $ \Psi_\gamma^t (Z_1,Z_2,\dots,Z_t) $  as follows:
  the set contains sequences $x_1,x_2,\dots,x_t$ 
whose probabilities $P(x_{1..t}|Z_{1..t}) $ are maximal and  sum to~$\gamma$.
This definition is not precise, since it is possible that the sum can not be made equal to $\gamma$ exactly.
That is why the formal definition of the  confidence set will use randomization.

For this purpose, we order all sequences $X_{1..t}$ according their conditional probabilities, 
in the decreasing order. That is, 
enumerate all sequences $x_{1..t}\in\X^n$  in such a way that  $(a_{1..t})\in\X^t$ has a smaller index than $(b_{1..t})\in\X^t$ if 
 either $P(a_{1..t}|Z_{1..t}) > P(b_{1..t}|Z_{1..t})$, or
$P(a_{1..t}|Z_{1..t}) = P(b_{1..t}|Z_{1..t})$
 and $(a_{1..t})$ is lexicographically less 
than $(b_{1..t})$.   Let $j$ be the integer for which 
$ \sum_{i=1}^{j-1} P(x^i_{1..t}| Z_{1..t}) \le \gamma$ and 
$ \sum_{i=1}^{j} P(x^i_{1..t}| Z_{1..t})> \gamma$.
If $\sum_{i=1}^{j-1} P(x^i_{1..t}| Z_{1..t})=\gamma$, 
then define $\Psi_\gamma^t (Z_{1..t})$ as the set $\{x^1_{1..t},\dots,x^{j-1}_{1..t}\}$. Otherwise, 
$ \Psi_\gamma^t (Z_{1..t})$ also contains $j-1$ first elements, and additionally  the element
$x^j_{1..t}$ with probability 
$( \gamma -  \sum_{i=1}^{j-1} P(x^i_{1..t}| Z_{1..t}) )/ P(x^j_{1..t}| Z_{1..t})$. 
(Note that this procedure is  commonly used in mathematical statistics for
making  the confidence level exactly  $\gamma$.) 
When talking about the sizes of the confidence sets we refer to their expected (with respect 
to the randomization) size.

Next,  we estimate the size of the described confidence~set. 
\begin{theorem}\label{th:1}
Let an (unknown) signal $X = X_1 X_2,\dots$ and its noisy version $Z= Z_1 Z_2,\dots$ be 
stationary ergodic processes with finite alphabets. 
Then, for every $\gamma \in (0,1)$, all $t\in\N$ and 
almost every  $Z_1,\dots,Z_t$
 the  confidence set $ \Psi_\gamma^t (Z_1,\dots,Z_t) $  contains the unknown
$(X_1,\dots,X_t)$  with probability $\gamma$:
\begin{equation}\label{t1a2}
 P\{ X_{1..t} \in \Psi_\gamma^t(Z_{1..t})\} = \gamma,
\end{equation}
while,
with probability 1,  the size of the set $ \Psi_\gamma^t (Z_1,\dots, Z_t)$
grows exponentially with the exponent rate 
that is equal to the conditional entropy:
 \begin{equation}\label{t1}
  \lim_{t \rightarrow \infty }   \frac{1}{t} \log \E | \Psi_\gamma^t (Z_1,\dots,Z_t)  | = h(X|Z)\ a.s.,
 \end{equation}
where the expectation is with respect to the randomization used in constructing the confidence sets.
\end{theorem}
\begin{proof} 
The proof of (\ref{t1a2}) immediately follows from  the construction of the set  $ \Psi_\gamma^t (Z_1 Z_2 ... Z_t)$.

The proof of (\ref{t1})  will be based on the Shannon-McMillan-Breiman theorem \cite{Cover:06,Gallager:68}, 
which for the conditional
entropy implies the following:
\begin{lemma}[Shannon-McMillan-Breiman]\label{l:smb}
  $\forall \varepsilon >0, \forall\delta > 0$, for almost all \\ $Z_1,Z_2,\dots$ there exists $n'$ such that 
  if $n > n'$  then 
\begin{equation}\label{smb}
   P\left\{ \left| -  \frac{1}{n} \log P(X_{1.._n}|Z_{1..n}) 
 - h(X|Z) \right| < \epsilon \right\} \ge 1-\delta. 
\end{equation}
\end{lemma}
Take any
$ \epsilon >0$ and any  $\delta>0$ such that
\begin{equation}\label{ecz}
1-\delta\ge\gamma.
\end{equation} 
 According to the lemma, for almost all $Z_1,Z_2,\dots$
there exists $n'$ such that (\ref{smb}) is valid if  $ n > n'$. Take any such  $n$  and rewrite (\ref{smb}) as follows:
\begin{equation}\label{eq:smb2}
  P \left\{ 2^{-n(h(X|Z) + \varepsilon)} \le P(X_{1..n} | Z_{1..n}) \le 2^{-n(h(X|Z) - \epsilon)} \right\} \ge 1-\delta.
\end{equation}
Thus, the probability of all   strings $x_1,\dots,x_n$ 
for which  we have $P(x_{1..n}| Z_{1..n})\ge2^{-n(h(X|Z) + \epsilon)}$ is at least $(1-\delta)$.
Taking into account (\ref{ecz}), we have 
$$  | \Psi_\gamma^t (Z_{1..n}) | \le \gamma / 2^{-n(h(X|Z) + \epsilon)},$$ so that
\begin{equation}\label{eq:kl}
{1\over n} \log | \Psi_\gamma^t (Z_{1..n}) |  \le  
h(X|Z) + \epsilon + O(1/n) 
\end{equation}
for $n>n'$.
Having taken into account that~(\ref{eq:kl}) holds for every small $\epsilon>0$ 
 we obtain  (\ref{t1}).
\end{proof}

\section{Optimality of the confidence set}
\begin{theorem}
Let an (unknown) signal $X = X_1 X_2,\dots$ and its noisy version $Z= Z_1 Z_2,\dots$ be 
stationary ergodic processes with finite alphabets $\X$ and $\Z$. 
Let $\Phi_\gamma^t (Z_{1..t})$,  be confidence sets, such 
that for some $\gamma\in(0,1)$ we have $P\left(X_{1..t}\in\Phi_\gamma^t (Z_{1..t})\right)\ge\gamma$  for all  $t\in\N$ and almost all  $Z_{1...t}\in\Z^t$.
Then, with probability~1,
\begin{equation}\label{t1a-}
\liminf_{t \rightarrow \infty }   \frac{1}{t} \log | \Phi_\gamma^t (Z_1,\dots,Z_t)  | \ge h(X|Z).
 \end{equation}
\end{theorem}
\begin{proof} 
The proof will  use the Shannon-McMillan-Breiman theorem~(\ref{eq:smb2}). 
As before,  we take any
$ \epsilon >0$ and fix  $\delta:=\gamma/2$. Then from some $n$ on we have~(\ref{eq:smb2}). 
Let $\Upsilon$ be a confidence set for  this $n$ and a certain $\gamma$. Define 
\begin{equation}\label{fi} 
 \Phi = \left\{ x_{1..n} : 2^{-n(h(X|Z) + \varepsilon)} \le\right. 
\left.P(x_{1..n} | Z_{1..n}) \le 2^{-n(h(X|Z) - \varepsilon) }\right\}.
\end{equation}
By definition, 
$ \sum_{x_{1..n} \in \Upsilon} P(x_{1..n} | Z_{1..n}) \ge \gamma. $
From this and~(\ref{eq:smb2})  we obtain 
$$ \sum_{x_{1..n} \in \Upsilon \cap 
\Phi} P(x_{1..n} | Z_{1..n}) \ge \gamma - \delta. 
$$
From this and~(\ref{fi}) we get
$$ |\Upsilon| \ge |\Upsilon \cap 
\Phi| \ge  (\gamma - \delta) 2^{ n(h(X|Z) - \epsilon) }. 
$$
Hence, 
$$
\liminf_{t \rightarrow \infty }   \frac{1}{n} \log | \Upsilon| \ge h(X|Z) - \varepsilon .
$$
Since this inequality 
 is true for    any  confidence set   $\Upsilon$  and any $\varepsilon > 0$, we obtain~(\ref{t1a-}).
\end{proof}

\section{Erasure channel with unknown statistics}
In this section we  consider the case when the channel statistics is unknown, but the channel
has a specific form: it is an erasure channel  for which probabilities to be erased are equal for all symbols. 
We show that the described above confidence set is asymptotically  optimal in this case, too. 
The point is that in this the conditional probabilities  $ P(X_{1..n}/ Z_{1..n}) $ are known, that is why
the construction of the previous section is directly applicable. 

The formal description of the considered model is as follows. 
We still assume that there is a known stationary ergodic  source  generating the signal $X_1,X_2,\dots$.
 The erasure channel is defined in two following  steps: first, 
there is a stationary ergodic process $\Theta$ generating letters from the  alphabet $\{ \Lambda, * \}$ and, second, 
 the noisy channel  is 
determined by  the following ``summation''  of the (uncorrupted) sequence 
$X_1,X_2,\dots$ and the noise sequence  $\Theta_1,\Theta_2,\dots$:
$$
 Z_i = \begin{cases} X_i& \text{ if } \Theta_i = \Lambda \cr
*  &
\text{ if }  \Theta_i = *.\end{cases}
$$


\begin{theorem}\label{th:er}
Let an (unknown) signal $X = X_1 X_2,\dots$ and $Z_1,Z_2,\dots$ be a stationary ergodic signal and its 
version corrupted by an unknown  stationary erasure channel. 
Then, for every $\gamma \in (0,1)$, all $t\in\N$ and 
almost every  $Z_1,\dots,Z_t$
 the (above described)   confidence set $\Psi_\gamma^t (Z_1,\dots,Z_t) $  contains the unknown
$(X_1,\dots,X_t)$  with probability $\gamma$:
\begin{equation}\label{t1a}
 P\{ X_{1..t} \in \Psi_\gamma^t(Z_{1..t})\} = \gamma,
\end{equation}
while,
with probability 1,  the size of the set $ \Psi_\gamma^t (Z_1,\dots, Z_t)$
grows exponentially with the exponent rate 
that is equal to the conditional entropy:
 \begin{equation}\label{t12}
  \lim_{t \rightarrow \infty }   \frac{1}{t} \log \E | \Psi_\gamma^t (Z_1,\dots,Z_t)  | = h(X|Z)\ a.s.,
 \end{equation}
where the expectation is with respect to the randomization used in constructing the confidence sets.
\end{theorem}
\begin{proof} 
It is enough to notice that, although the erasure channel is not known, the probabilities $P(X_{1..n}|Z_{1..n})$ are known. 
Therefore, the proof of this theorem is identical to that of Theorem~\ref{th:1}.
\end{proof}

\section{Discussion}
To the best of our knowledge, the problem of constructing a   confidence set 
 for the unknown signal was not considered before, which is why there are many quite natural 
and obvious extensions and generalizations of the present work. 
First, it is  interesting  to consider this problem
for certain  specific classes of distributions of the signal and  noise, such as  i.i.d.\ and Markov sources.
For these classes of sources it should be possible to obtain rates of convergence in those
statements that in this work  are only asymptotic, for example in~(\ref{t1}).
 
Second, a natural question is to find a construction of the confidence set for the cases where 
the signal is multi-dimensional. This is particularly important for applications, many  of which are  concerned with 
denoising such objects as photographs or video fragments. 
Another interesting generalization is the case where the alphabets 
are (subsets of), for example,  the Euclidean space. This generalization can be also  interesting from the practical point of view. 
Finally, the case where statistics of the noise and/or signal are unknown is obviously of  great theoretical and practical
interest.

 \subsection*{Acknowledgments}
Boris Ryabko was partially supported  by the  Russian Foundation for Basic Research
 (grant no. 09-07-00005). Daniil Ryabko was partially supported  by  the French Ministry of Higher Education and Research, 
Nord-Pas de Calais Regional Council and FEDER through CPER 2007-2013, 
  ANR projects  EXPLO-RA (ANR-08-COSI-004) and Lampada (ANR-09-EMER-007), and by Pascal-2.
Some of these results were reported at ISIT 2011 \cite{rr11}.

\end{document}